# THE PRINCIPLES OF HUMANISM FOR MANETS

Md. Amir Khusru Akhtar[1] and G. Sahoo[2]

[1]Department of CSE, Cambridge Institute of Technology, Ranchi, Jharkhand, India
[2]Department of IT, Birla Institute of Technology, Mesra, Ranchi, India

*ABSTRACT*

*The proposed humanistic approach mapped the human character and behavior into a device[1] to evade the bondages of implementation and surely succeed as we live. Human societies are the complex and most organized networks, in which many communities having different cultural livelihood. The formation of communities within a society and the way of associations can be mapped to MANET. In this work we have presented the principles of humanism for MANETs. The proposed approach is not only robust and secure but it certainly meets the existing challenges (such as name resolution, address allocation and authentication). Its object oriented design defines a service in terms of Arts, Culture, and Machine. An 'Art' is the smallest unit of work (defined as an interface), the 'Culture' is the integration and implementation of one or more Arts (defined as a class) and finally the 'Machine' which is an instance of a Culture that defines a running service. The grouping of all communicable Machines having the same Culture forms a 'Community'. We have used the term 'Society' for a MANET having one or more communities and modeled using the humanistic approach. The proposed approach is compared with GloMoSim and the implementation of file transfer service is presented using the said approach. Our approach is better in terms of implementation of the basic services, security, reliability, throughput, extensibility, scalability etc.*

*ABBREVIATIONS*

*Community Table (CT), Society Table (ST), Machine Culture (MC), Machine Identification (MID), Community Identification (CID), Service Initiator (SI), Machine Culture Start (MCSTART), Machine Culture Join (MCJOIN)*

## 1. INTRODUCTION

As one can access the MANETs essence in the society, its immense use, the prolonged scatterration and the future requirement the foremost and futuristic thoughts can be further redefined and exaggerated from the needs of the acclimating society. The work which has been done so far has been very curative but it is high time to amend the ways for an energetic means to create a new platform for novel research. The bondages of this study should finally pave a new way, for further development and create a new awareness amongst the researchers. By the involvement of the existing legacy and incorporating the necessary demand new eras will surely arrive.

---

[1] It represents electronic equipment such as Computer, PDA, Mobile phones etc.

To realize the proposed approach first of all we have to understand the formation of human communities and societies. A human community is formed to meet some objective(s). We have one or more communities within a society and a person is attached with one or more communities to fulfil its needs. These concepts of association and cooperation can be mapped to MANET, because in a network also nodes are running one or more services to fulfil its needs. We have used the term 'Community' to represent a group of communicable machines running a service having the same culture. Further, a MANET needs several services that's why we need to create several communities to provide more than one services such as for file service, name service. These services can be named as file sharing Community, name service Community, etc.

The proposed humanistic approach mapped the human attributes and behavior into a device to evade the bondages of implementation and surely succeed as we live. Human network is the complex topology that still survives even in the case of known/unknown incidents. Our proposed approach assumes a node as a person that belongs to a society and further it is associated with one or more communities. That's why, this anatomy is required.

This approach employs 'Art' the smallest unit of work (defined as an interface), the 'Culture' is the integration and implementation of one or more Arts (defined as a class) and finally the 'Machine' that is an instance of a Culture and define a running service. A node must run a Machine to avail a service. The grouping of Machines forms a 'Community' having the same Culture, and provides a service within the network. Nodes can start a service and intimate this information to the network by sending a MCSTART packet. Nodes that want to join and use the service send its consent using MCJOIN packet. The grouping of multiple different Communities forms a Society.

The Arts can be implemented in terms of fine grain or coarse grain level. Defining Art at fine grain level creates complexity in design but gives high customization. The object oriented implementation surely gives a robust, secured, extensible and reliable platform to execute an application in the MANET that meets all the challenges that we have.

The rest of the paper is organized as follows. Section 2 presents the assumption and basic primitives. Section 3 briefs the related work. Section 4 presents the proposed work. Section 5 gives an analysis of the proposed approach. A secure and energy saving solution is proposed in section 6. Section 7 shows the implementation of FTP using humanistic approach. Section 8 highlights the conclusion and ongoing work.

## 2. ASSUMPTIONS AND BASIC PRIMITIVES

This section outlines the assumptions and basic primitives that we make regarding our proposed approach.

### 2.1. Elementary terminologies

We have used following terms in our work

**Art (A):** it denotes the smallest unit of work module needed to define a Culture and can be implemented at fine grain or coarse grain level.
**Culture (C):** it denotes the integration and implementation of Arts that is needed to define a service.
**Machine (M):** it denotes an instance of a Culture that defines a running service and it should be a member of a Community.

**Community (C):** it denotes a group of communicable Machines having the same Culture that provide a service within the Society.

**Society (S):** it denotes the MANET consisting of one or more communities and modeled using humanistic approach.

### 2.2. Human/device common basic needs

To design a MANET using our proposed humanistic approach first we have to analyze the human and device basic needs to realize the benefits of association and to identify the associated setback. We have defined the common basic needs given in Table 1, to understand the fact and to establish a mapping between human and device.

Table 1. Basic Needs

| Human Needs | Device Needs | Art |
|---|---|---|
| Food | Energy | Energy saving Arts can be defined using the existing algorithms that save battery life and bandwidth. |
| Cloth | Privacy | Privacy Arts can be defined by involving existing techniques which fulfil privacy needs. |
| House | Security | Security Arts can be defined using existing security algorithms. |

### 2.3. Need of association

The association is an agreement between groups of individuals/nodes that form a community to accomplish a purpose. Peoples are living in a society in which they have different communities for different needs. We have defined these associations because in a Society/MANET nodes have common objectives.

#### 2.3.1. Communities

Community means a particular 'class' of peoples having the same objective. The human community formation concept can be mapped to MANET because in a network a set of nodes work together to run a service to fulfil its need.

#### 2.3.2. Society

A 'Society' refers to all communities of people associated together to fulfil diverse needs. We have used the term 'Society' for a MANET having one or more communities and modeled using the humanistic approach to provide different services in a network.

## 3. RELATED WORKS

The work which has been done so far has been very curative but it is high time to amend the ways for an energetic means to create a new platform for more research. The bondages of this study should finally pave a new way, for further development and create a new awareness amongst the researchers. By the involvement of the existing legacy and incorporating the necessary demand new eras will surely arrive.

To our knowledge, there is no such design that we have proposed and no previously published work that meets all the challenges that we have in Mobile ad hoc networks. Lots of work has been proposed to handle modification of routing information's [1-5], to enforce cooperation [6, 21-23] but they have serious limitations in terms of routing overhead and attacks. Work to be done in the area of Address allocation [7], name resolution [8-9] shows that the implementation of these services with the current design of MANET creates complexity and heavy network overhead.

## 4. PROPOSED WORK

MANETs are certainly the next generation face that announces the game, winning which we obtain lave. The known/unknown obstacles resist the implementation, and by circumventing we are able to utilize the resource efficiently. The aggregated/generalized form of MANET can further specialize to esteem the demand that is why we are going to restructure the same and prevail. Human nature would really act into the device to entrust and make it robust and secure like the same.

### 4.1. Overview

Our object-oriented design defines a service in terms of Arts, Culture, and Machine. This approach employs 'Art' the smallest unit of work (defined as an interface), the 'Culture' is the integration and implementation of one or more Arts (defined as a class) and finally the Machine which is an instance of a Culture and defines a running service. A node must run a Machine to avail a service. The Machine defines a running service in a node, and the grouping of Machine having the same Culture forms a 'Community' defined in section 4.4.4.

We can provide the basic services in a MANET by instantiating a Machine after fulfilling the basic requirements for some services (for ex. Internet gateway point for name service). A SI node broadcast MCSTART packet in the Society/MANET to intimate that a service is being started. To join the service a node after receiving the MCSTART must send a MCJOIN packet to become a member of the Community or to register itself into an existing Community. The lists of interested nodes that have replied by sending MCJOIN message are listed by the SI to form a Community.

We have used community tables in each Community to stores CID, MID and path. Packets are routed on the basis of the path defined in the community table. During transmission if a node doesn't find a route in the community table then it broadcasts a RREQ that is forwarded from node to node either node is a member or if in is not a member it handle it as friend RREQ. When the RREQ packet reaches to Community members, the members reply the source by sending the community table. This model defines levels of abstraction to handle malicious or selfish attacks defined in section 4.5. Hence, the proposed approach certainly gives us the platform upon which we can run and use any application in MANET. Figure 1 shows a MANET which is derived from human attributes and behavior.

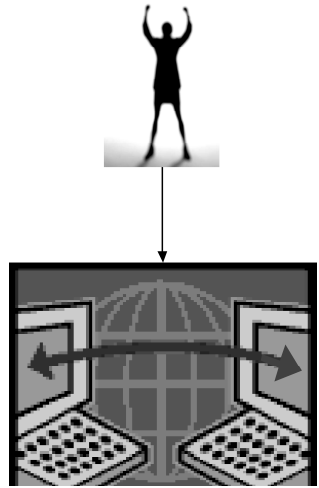

Figure 1. Humanistic MANET

Humanistic approach defines the concept that in a society we have different community and each community is for different purposes/services. Similarly in a network we can create many communities with the same set of nodes to run many services. For example, a node transfers a file using Community 1, and logged on onto a remote node using Community 2.

### 4.2. Implementation Diagram

A Machine is an instance of a Culture and a Culture implements one or more Arts to define a service. Figure 2 shows the implementation diagram in which we have defined a Machine using Arts and Cultures.

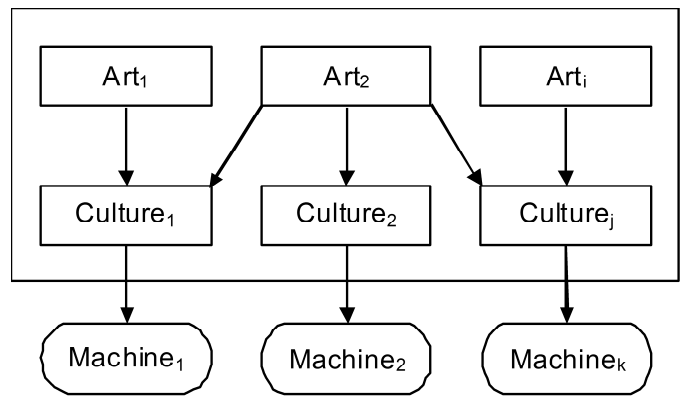

Figure 2. Diagram showing interface (Arts), Class (Cultures), and object (Machines) in which a Culture is defined using Arts and finally instantiated to form a Machine

where
    Arts ranges from $(1, 2, …, i)$
    Cultures ranges from $(1, 2, …, j)$
    Machines ranges from $(1, 2, …, k)$

## 4.3. Types of Society/MANET

This section presents two types of Society/MANET.

### 4.3.1. Single Community MANET

A 'single Community Society' is a MANET in which only one service is running such as file service, and it is created in the same way as defined in section 5.2. Figure 3 shows a single Community MANET.

### 4.3.2. Multiple Communities MANET

A 'multiple Communities Society' is defined as a MANET in which more than one services are running for example file service, name service etc. It is created in the same way as defined in section 5.2. Figure 4 shows a multiple Community MANET. We have used M (MC, MID) notation to denote a Machine

where

>    M is the Machine
>    MC is the Machine Culture
>    MID is the Machine Identity within the Culture

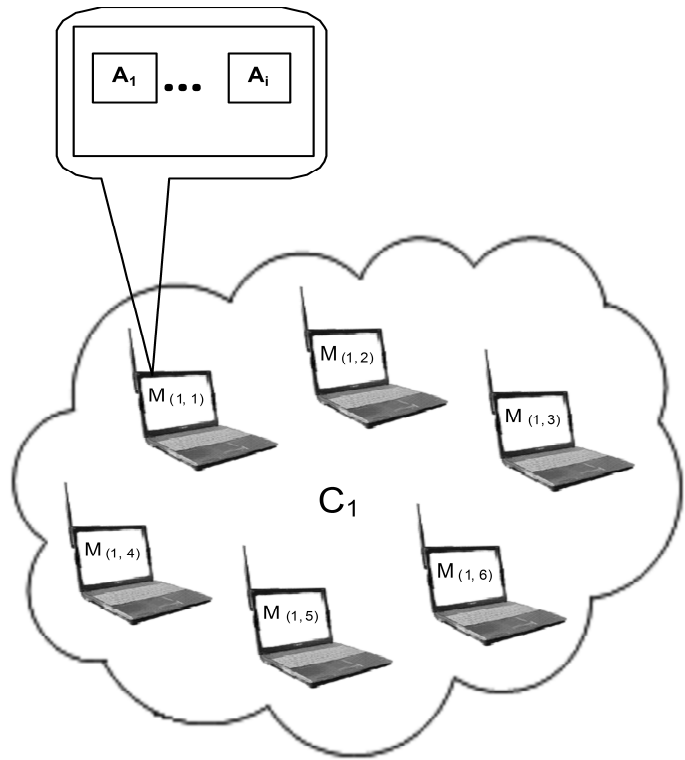

Figure 3. Single Community MANET

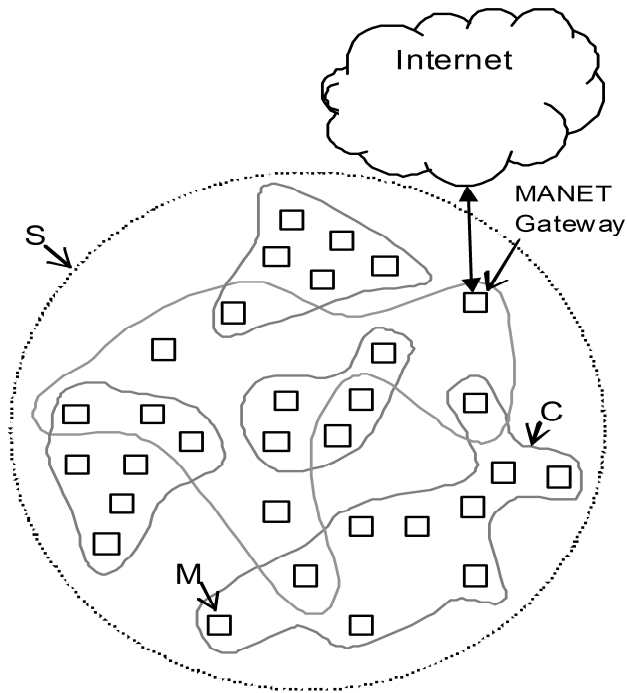

Figure 4. A Society consisting of 5 communities

## 4.4. Description of the components

The components of the proposed approach are as follows

### 4.4.1. Arts Component

It defines the smallest work module(s) that is used to define a service and implemented in Culture. Art(s) can be defined in terms of fine grain or coarse grain level. Defining Art at fine grain level creates complexity in design but gives high customization. The fine grain Art(s) can be defined at the function level (such as RREQ, RREP, RERR, HELLO etc.), similarly coarse grain Art(s) can be defined at the protocol level (such as Bellman-Ford, DSDV, FSR, OSPF, WRP, LAR, DSR, AODV) [10-17]. The Arts are defined in terms of interfaces that contain final variables and abstract methods defined in Table 2.

Table 2. Art Interface

| Interface Art1 |
|---|
| final variables; |
| abstract methods (); |

### 4.4.2. Culture Component

We have used Culture to integrate one or more Arts and defined in terms of a class as given in Table 3. Cultures implements one or more Arts and instantiated to form a Machine.

Table 3. Culture Class

| Class Culture1  implements  Art1, Art2, …,  Arti |
|---|
| Culture variables; |
| // implementation of methods declared in Arts interfaces<br>abstract methods ();<br>// implementation of Culture methods<br>Culture methods (); |

### 4.4.3. Machine Component and packets

A Machine is an instance of a Culture that defines a running service. To regularize its service a machine uses various functions that are defined in Culture methods (), they are as follows.

    **MCSTART**: Packet broadcasted to the Society/MANET to intimate that a service has been started.
    **MCJOIN**: Packet sends by interested Society/MANET member in response of MCSTART to join the Community.

### 4.4.4. Community Component

Community means the grouping of two or more Machines having the same Culture and communicable either directly or indirectly. Each Community maintains a community table listing all the members of the Community. A Community network is shown in figure 5 through which community tables can be defined. We have shown a community table for node N1 in Table 4, which is defined using figure 5.

Table 4. Community Table

| MID | CID | PATH |
|---|---|---|
| N2 | C1 | N1-N2 |
| N3 | C1 | N1-N3 |
| N4 | C1 | N1-N2-N4 |

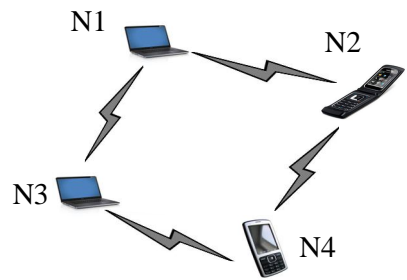

Figure 5. Community network

### 4.4.5. Society Component

All Communities finally constitute a Society that is defined for diverse goals. The Society Table is defined in Table 5.

Table 5. Society Table

| CID | MC |
|---|---|
| C1 | File service |
| C2 | Name Service |

### 4.5. Levels of Abstraction

The communication between Machines (object) hides the internal details. The description of data at the object level is in a format independent of its Arts and Culture representation. Our work defines various levels of abstraction to minimize the complexity and protect the system from selfish and malicious threats as is given in figure 6.

Our object oriented design confirms that only defined operations can be executed with nodes because data's are wrapped by operations, which protects our network from malicious and selfish attacks.

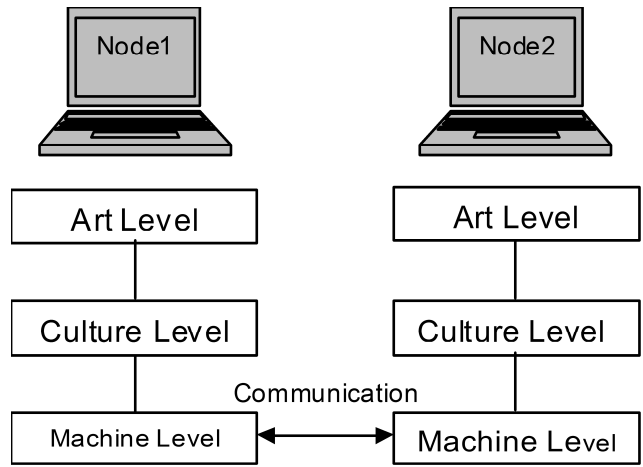

Figure 6. Levels of Abstraction

## 5. ANALYSIS OF THE PROPOSED APPROACH

In this section we have described the comparison between proposed structure and GloMoSim and the implementation of basic services using the proposed approach.

### 5.1. Comparison between proposed Structure and GloMoSim Layers

The GloMoSim Layers [18-19] and the proposed structure are given in Table 6 and Table 7 respectively. Through these tables we have compared our proposed structure with GloMoSim structure.

Table 6. Layered Structure of GloMoSim

| Layer | Protocol |
|---|---|
| Physical (Radio Propagation) | Free space, Two-Ray |
| Data Link (MAC) | CSMA, MACA, TSMA, 802.11 |
| Network (Routing) | Bellman-Ford, FSR, OSPF, DSR, WRP, LAR, AODV |
| Transport | TCP, UDP |
| Application | Telnet, FTP |

International Journal of Computer Science & Information Technology (IJCSIT) Vol 5, No 5, October 2013

Table 7. Proposed Structure

| Arts | | | | |
|---|---|---|---|---|
| Physical (Radio Propagation) | Data Link (MAC) | Network (Routing) | Transport | Application |
| Free space Two-Ray | CSMA MACA TSMA 802.11 | Bellman-Ford OSPF FSR DSR WRP LAR AODV | TCP UDP | Telnet FTP |

| Cultures | | |
|---|---|---|
| Culture 1 | Culture 2 | Culture 3 |
| Free space CSMA Bellman-Ford TCP FTP | Two-Ray 802.11 DSR TCP CBR | Two-Ray CSMA AODV TCP Telnet |

### 5.2. Implementation of basic services

The Aleksi Penttinen [20] defined various challenges and services such as Address allocation, name resolution, authentication etc. Work to be done in the area of Address allocation, name resolution [7, 9] shows that the implementation of these services with the current design of MANET creates complexity and heavy network overhead. These services can be easily implemented using the proposed approach without such problems.

To explain the implementation of these services first we relate our concept with the creation of community in human. For example to create a community or club for some specific activity, two or more persons joined and make a community and then they add interested members and avail the services of the community.

#### 5.2.1. Working Diagram

Our proposed object oriented design is simple and extensible than the traditional MANET design. To deploy a new service we simply create a list of Arts needed for the service (such as name service) and implement these Arts into a Culture to instantiate a Machine. After that, the SI node broadcast MCSTART packet in the Society/MANET to intimate that a service is being started. To join the service a node after receiving the MCSTART must send a MCJOIN packet to become a member of the Community. The lists of interested nodes that have replied by sending MCJOIN message are listed by the SI to form a Community. The working diagram for implementing basic services is given in figure 7. However, for name service any of the nodes should be connected to the internet to provide a gateway for the network to resolve the name.

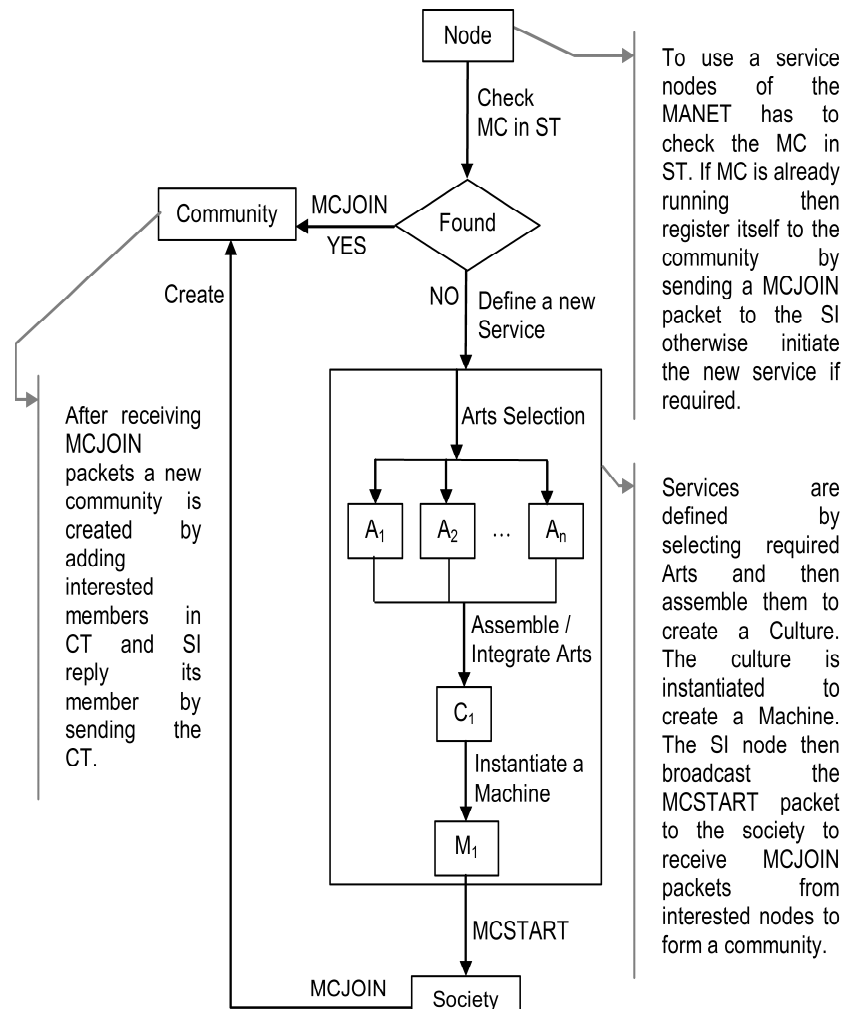

Figure 7. Working model for initiating Basic services

### 5.2.2. Friend packets

If a node does not find a route in the community table it sends friend packets to its reachable host to forward its packet to its Community members. After that the receiving host relay packets to the exact Community member.

### 5.2.3. Overhead

In our approach we were sending only one broadcast message at the time when a service is started to create a Community. After that we simply multicast messages within the Community to regularize and avail the services, thus it degrades the message overhead in the MANET.

## 6. PROPOSED SECURE AND ENERGY SAVING DESIGN

On the basis of the proposed structure as defined in section 5.1 a secure and energy saving solution can be defined. To make a secure and energy saving protocol all existing solutions are classified in terms of Basic and Add-on protocols. Basic protocols (such as DSR [16], AODV [17] etc.) are suitable for cooperative or closed environment such as in military or police exercises, disaster relief operations, and mine site operations. Add-on protocols are used for non cooperative or open MANET, in which nodes agreed to participate in network activities but for saving itself they misbehave. The existing methods (such as Watchdog and Pathrater [6], CONFIDANT [21-22], CORE [23], etc.) can be used with the Basic protocol to enforce cooperation. Closed and Open MANET is well defined by Akhtar and Sahoo [24]. The best suitable network environment and corresponding protocol are given in Table 8.

Table 8. Suitable environment to launch correct protocols

| Network Environment | Protocol |
| --- | --- |
| Closed MANET (It is a cooperative network having minimum threats) | P1 |
| Open MANET (It is an uncooperative network having highest threats) | P2 |

On the basis of the environment, the protocol components can be defined. Table 9 shows the protocol components in which protocol (P1) uses only basic protocol because it is used in a cooperative environment while the protocol (P2) uses basic plus add-on protocol because it is used in non cooperative environment.

Table 9. Protocol Components

| Protocol Name | Basic Protocol | Add-on Protocol |
| --- | --- | --- |
| P1 | DSR | NO |
| P2 | AODV | CORE or CONFIDANT |

On the basis of network environment correct protocol can be launched which ensures minimum battery consumption. Hence, the proposed approach enhances cooperation in MANET because energy is the main cause of misbehave.

## 7. CASE STUDY: FILE TRANSFER SERVICE

In this section we have proposed how File transfer service can be implemented using the proposed approach. The protocol stack (Arts) needed to implement File transfer service is defined in Table 10.

Table 10. Protocol stack (Arts)

| Culture F |
| --- |
| Free space |
| CSMA |
| DSDV |
| TCP |
| FTP |

The working model for the implementation of file transfer service is similar to the implementation of basic services as defined in figure 7. This approach minimizes unnecessary broadcast because our approach divides a MANET in terms of Community while the traditional methods perform regular updates of its routing tables in the entire network. We have created Community, in which network participation varies from node to node. Only interested members are responsible for the network activities and thereby exclude the participation of idle nodes. Hence, the proposed approach saves battery power and bandwidth as well as it is secure because it allows Machine (object) communication.

## 8. CONCLUSION AND ONGOING WORK

In this paper we have presented the principles of humanism for MANETs which mapped the human attributes and behavior into a device to evade the bondages of implementation. Our approach is not only robust and secure but it certainly meets the existing challenges of the MANET, because of its object oriented design. We have discussed the basic services and presented how these services can be implemented using the said approach. In the case study we have proposed the File transfer service using our humanistic approach and it justifies, that it is better than the traditional implementation in terms of overhead and attacks. This paper presents initial work on the humanistic MANET design. The long-term goal of this research project is to implement all the know-how in the MANET and to design a humanistic network simulator that fulfills the basics of the MANET.

## REFERENCES

[1] K. Sanzgiri, B. Dahill, B. Levine, C. Shields, and E. Belding-Royer, "A secure routing protocol for ad hoc networks," in the 10th IEEE International Conference on Network Protocols (ICNP), November 2002.
[2] M. G. Zapata and N. Asokan, SAODV "Securing ad-hoc routing protocols", in the 2002 ACM Workshop on Wireless Security (WiSe 2002), September 2002, pp.1-10.
[3] Y. Hu, A. Perrig, and D. Johnson. Ariadne: "A Secure On-Demand Routing Protocol for Ad Hoc Networks", In Proceedings of the Eighth Annual International Conference on Mobile Computing and Networking, September 2002, pages 12-23.
[4] Johnson, and A. Perrig. SEAD: "Secure Efficient Distance Vector Routing in Mobile Wireless Ad Hoc Networks', in Fourth IEEE Workshop on Mobile Computing Systems and Applications, June 2002, pages 3-13.
[5] P. Papadimitratos, Z. Haas and P. Samar. "The Secure Routing Protocol (SRP) for Ad Hoc Networks", internet-Draft, draft-papadimitratos-securerouting- protocol-00.txt, December 2002.
[6] S. Marti, T. J. Giuli, K. Lai, and M. Baker, "Mitigating routing misbehavior in mobile Ad hoc networks," in Proceedings of the 6th annual international conference on Mobile computing and networking (ACM MobiCom 2000), pp. 255–265, New York, NY, USA, 2000.
[7] Xiaonan Wang, Yi Mu, "A secure IPv6 address configuration scheme for a MANET", Security and Communication Networks, 2012, doi: 10.1002/sec.611
[8] M. Nazeeruddin, G.P. Parr, B.W. Scotney, "An efficient and robust name resolution protocol for dynamic MANETs", Ad Hoc Networks, Volume 8, Issue 8, November 2010, Pages 842-856
[9] Mohammad Masdari, Mehdi Maleknasab, Moazam Bidaki, "A survey and taxonomy of name systems in mobile ad hoc networks", Journal of Network and Computer Applications, Volume 35, Issue 5, September 2012, Pages 1493–1507
[10] RE Bellman, "On a routing problem", Quarterly of Applied Mathematics, 16 (1958), pp. 87–90.
[11] Y. Hu, D. C.E. Perkins & P. Bhagwat, "Highly Dynamic Destination-Sequenced Distance Vector Routing (DSDV) for Mobile Computers", Proceedings of ACM SIGCOMM'94, London, UK, Sep. 1994, pp. 234-244.

[12] Guangyu Pei, M. Gerla, and TsuWei Chen, "Fisheye state routing: a routing scheme for ad hoc wireless networks", Communications 2000 ICC 2000, 2000 IEEE International1Conference on, 1:70–74 vol.1, 2000.

[13] Moy, J., "The OSPF Specification", RFC 1131, October 1989.

[14] Shree Murthy and J. J. Garcia-Luna-Aceves, "An Efficient Routing Protocol for Wireless Networks", Mobile Networks and Applications, 1(2):183–197, 1996.

[15] Y. B. Ko, and N. H. Vaidya, "Location-Aided Routing (LAR) in Mobile Ad hoc Networks", Proceedings of the 4th annual ACM/IEEE international conference on Mobile computing and networking, 1998.

[16] Johnson, D. Maltz, and Y.C. Hu, "The Dynamic Source Routing Protocol for Mobile Ad Hoc Networks (DSR)," IEEE Internet Draft, Apr. 2003.

[17] C. Perkins, E. B. Royer, S. Das, "Ad hoc On-Demand Distance Vector (AODV) Routing", IETF Internet Draft, 2003.

[18] http://pcl.cs.ucla.edu/projects/glomosim/

[19] Lokesh Bajaj, Mineo Takai, Rajat Ahuja, Ken Tang, Rajive Bagrodia, and Mario Gerla, "Glomosim: A scalable network simulation environment", UCLA Computer Science Department Technical Report 990027 (1999): 213.

[20] Aleksi Penttinen, "Research on Ad Hoc Networking: Current Activity and Future Directions", Networking Laboratory, Helsinki University of Technology, Finland. See also http://citeseer.nj.necm.com/533517.html.

[21] S. Buchegger and J. Y. Le-Boudec, "Performance analysis of the CONFIDANT protocol (cooperation of nodes: Fairness in dynamic ad-hoc networks)", In Proceedings of MobiHOC'02, June 2002.

[22] S. Buchegger and J. Y. Le-Boudec, "Nodes bearing grudges: Towards routing security, fairness, and robustness in mobile ad hoc networks", In Proceedings of EUROMICRO- PDP'02, 2002.

[23] P. Michiardi and R. Molva, "CORE: A Collaborative Reputation Mechanism to enforce cooperation in Mobile Ad-hoc networks", in CMS'2002, Communication and Multimedia Security 2002 Conference, September 26-27, 2002, Portoroz, Slovenia / Also published in the book : Advanced Communications and Multimedia Security /Borka Jerman-Blazic & Tomaz Klobucar, editors, Kluwer Academic Publishers, ISBN 1-4020-7206-6, August 2002 , 320 pp, August 2002.

[24] Md. Amir Khusru Akhtar, G. Sahoo, "A Novel Methodology for Securing Adhoc Network by Friendly Group Model", The Fourth International Conference on Networks & Communications (NetCoM) Chennai, Lecture Notes in Electrical Engineering, Springer, Series ISSN 1876-1100, Series Volume 131, pp 23-35, Jan 2013. DOI - 10.1007/978-1-4614-6154-8_3.

## Authors

**Md. Amir Khusru Akhtar** received his M.Tech. in Computer Science. & Engg. from Birla Institute of Technology, Mesra, Ranchi, India in the year 2009 and is pursuing PhD in the area of Mobile Adhoc Network from Birla Institute of Technology, Mesra, Ranchi, India. Currently, he is working as an Assistant Professor in the Department of Computer Science and Engineering, CIT, Tatisilwai, Ranchi, Jharkhand, India. His research interest includes mobile adhoc network, network security, parallel and distributed computing and cloud computing.

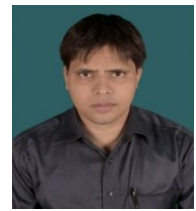

**G. Sahoo** received his M.Sc. in Mathematics from Utkal University in the year 1980 and PhD in the Area of Computational Mathematics from Indian Institute of Technology, Kharagpur in the year 1987. He has been associated with Birla Institute of Technology, Mesra, Ranchi, India since 1988, and currently, he is working as a Professor and Head in the Department of Information Technology. His research interest includes theoretical computer science, parallel and distributed computing, cloud computing, evolutionary computing, information security, image processing and pattern recognition.

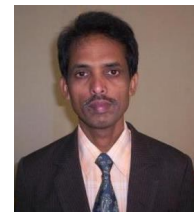